\documentclass[conference]{IEEEtran}
\IEEEoverridecommandlockouts
\usepackage{cite}
\usepackage{amsmath,amssymb,amsfonts}
\usepackage{graphicx}
\usepackage{textcomp}
\usepackage{xcolor}
\usepackage[numbers,square]{natbib}
\usepackage{algorithm}
\usepackage{algorithmicx}
\usepackage{algpseudocode}
\usepackage{threeparttable}
\usepackage{subfig}
\usepackage{tcolorbox}
\usepackage{multirow}
\usepackage{url}
\usepackage{makecell}
\usepackage{hyperref}

\def\BibTeX{{\rm B\kern-.05em{\sc i\kern-.025em b}\kern-.08em
    T\kern-.1667em\lower.7ex\hbox{E}\kern-.125emX}}
\begin{document}

\title{An Empirical Study on the Distance Metric in Guiding Directed Grey-box Fuzzing
}

\author{

\IEEEauthorblockN{Tingke Wen, Yuwei Li\thanks{Tingke Wen and Yuwei Li are the co-first authors.}, Lu Zhang, Huimin Ma, Zulie Pan\thanks{Zulie Pan is the corresponding author.}}
\IEEEauthorblockA{College of Electronic Engineering, National University of Defense Technology, Hefei, China}
\IEEEauthorblockA{\{wentk, liyuwei, zhanglu\_zl, mahuimin17, panzulie17\}@nudt.edu.cn}

}

\maketitle
\thispagestyle{plain}

\begin{abstract}

Directed grey-box fuzzing (DGF) aims to discover vulnerabilities in specific code areas efficiently. 
Distance metric, which is used to measure the quality of seed in DGF, is a crucial factor in affecting the fuzzing performance.
Despite distance metrics being widely applied in existing DGF frameworks, it remains opaque about how different distance metrics guide the fuzzing process and affect the fuzzing result in practice.
In this paper, we conduct the first empirical study to explore how different distance metrics perform in guiding DGFs. 
Specifically, we systematically discuss different distance metrics in the aspect of calculation method and granularity. 
Then, we implement different distance metrics based on AFLGo.
On this basis, we conduct comprehensive experiments to evaluate the performance of these distance metrics on the benchmarks widely used in existing DGF-related work. 
The experimental results demonstrate the following insights.
First, the difference among different distance metrics with varying methods of calculation and granularities is not significant.
Second, the distance metrics may not be effective in describing the difficulty of triggering the target vulnerability.
In addition, by scrutinizing the quality of testcases, our research highlights the inherent limitation of existing mutation strategies in generating high-quality testcases, calling for designing effective mutation strategies for directed fuzzing.
We open-source the implementation code and experiment dataset to facilitate future research in DGF.

\end{abstract}


\section{Introduction}\label{sec:intro}

Fuzzing is an efficient vulnerability discovery technique that has been widely applied in the real world~\cite{fuzzit, cifuzz, ossfuzz, syzkaller}.
Regarding different aims, fuzzing can be classified into two categories: coverage-based fuzzing and directed fuzzing.
Directed fuzzing aims to find vulnerabilities in the specified code target, while coverage-based fuzzing aims to test all code.
As vulnerable code only takes a tiny part of the whole program and achieving high code coverage is challenging,  directed fuzzing is more efficient than coverage-based fuzzing in detecting vulnerabilities located in the specified code area, especially for testing large-scale software~\cite{G-Fuzz}.

Early directed fuzzing techniques are primarily white-box, relying on symbolic execution~\cite{katch, wbfuzz01bohme, wbfuzz02icse, wbfuzz03sas}.
However, the directed white-box fuzzing techniques usually suffer from scalability and compatibility issues~\cite{Wang2020SoKTP}.
To address these challenges, directed grey-box fuzzing has been proposed~\cite{AFLGo, Windranger, hawkeye, CAFL,1dVul, UAFuzz, SeiveFuzz, SelectFuzz, TargetFuzz, KCFuzz}.
This approach is considered more practical than directed white-box fuzzing. 
Generally, distance-guided DGF first measures the distance between the execution trace of a seed and the location of the target code. 
Then, a seed with a shorter distance will be prioritized and assigned more energy in the mutation stage. 
The philosophy behind this is that if the path that the current seed executes is closer to the target code, then more mutations on that seed are more likely to generate expected testcases that reach the target code~\cite{Wang2020SoKTP}. 
Consequently, the target code will be adequately tested, and potential vulnerabilities will be discovered.

However, many open questions about the effectiveness of distance metrics remain undiscussed, despite its crucial and prevalent role in existing DGF frameworks.

\textbf{How do different distance calculation methods perform in practice?}
Although distance metric is widely used in existing DGF methods, it has different calculation methods such as harmonic mean distance, arithmetic mean distance, and the closest target distance, etc.
For instance, AFLGo~\cite{AFLGo} employs the harmonic mean instead of the arithmetic mean when calculating distances between basic blocks. 
The authors claim that harmonic mean can better differentiate between a node closer to one target but farther from another, versus a node equidistant from both targets. 
Which kind of distance performs better is still unknown and needs to be explored.

\textbf{Does the distance with finer granularity perform better in guiding directed fuzzing?}
There are different granularities in existing DGF distance calculating, such as function-level, basic-block-level, and approximated basic-block-level distance.
Compared to function-level distance, basic-block-level distance has finer granularity.
Intuitively, the distance with finer granularity can provide more precise feedback information and might be better.
However, obtaining fine granularity distance is not easy and may cause more overhead.
For instance, calculating precise function-level distance requires indirect call analysis, which is still a complex problem that has not been fully solved in the program analysis field~\cite{lu2019does}.
Additionally,  if two basic blocks are located in different functions, the basic-block-level distance of them should be calculated based on an inter-procedural control flow graph.
The existing DGFs usually use distance with different granularities. 
AFLGo calculates approximated basic-block-level distance mainly based on function-level distance.
Hawkeye augments distance calculation by recognizing the immediate call patterns between functions and achieves higher precision~\cite{hawkeye}. 
It is unknown whether the distance with finer granularity performs better in guiding directed fuzzing, and it should be evaluated.

\textbf{Do the seeds with closer distance have more contribution in generating PoC?} 
During the directed fuzzing process, the seeds with closer distance tend to be assigned with more power, which means that the closer seeds will be mutated more times to generate new testcases.
Note that the ultimate goal of directed fuzzing is to generate PoC (Proof-of-Concept), which can trigger the target. 
It is still unclear whether the closer seeds have more capabilities to achieve the goal.
Therefore, to evaluate the effectiveness of distance metrics, it is necessary to measure the contribution of closer seeds in generating PoC.

To answer the above questions, we conduct an empirical study on the effectiveness of different distance metrics in guiding directed fuzzing.
Our comprehensive discussion delves into the intricacies of typical distance metrics, covering various calculation methods and granularities. 
In specific, the typical distance calculation methods contain harmonic mean distance, arithmetic mean distance, and the closest target distance.
The typical distance granularity contains function-level distance, approximated basic-block-level distance, and basic-block-level distance.
Subsequently, we put our research into practice by implementing the corresponding DGFs with different distance metrics, using AFLGo as a basis, and rigorously evaluating them on real-world benchmarks.
Finally, we explore the validity of the closer distance in generating PoC that can trigger the target code quantitively and analyze the effectiveness of the existing DGF mutation strategy.
The highlighted findings in our work are as follows.

\begin{itemize}

\item In aspect of the distance calculation method, arithmetic mean distance performs better than harmonic mean and closest target distance.
Nevertheless, the advantage of arithmetic mean distance is not very obvious.
Regarding distance granularity, function-level distance, as the coarsest granularity distance, does not perform better than approximated basic-block-level distance and basic-block-level distance.
As a whole, the difference among distance metrics in terms of different calculation methods and granularities is not significant.

\item The distance metrics we examined encounter a bottleneck in selecting high-quality seeds and guiding directed fuzzing. 
The reason may be their deviation from accurately describing the difficulty of approaching and triggering the target vulnerability.

\item Our findings indicate that the current DFG mutation strategy may not be efficient, with the distance of testcases only decreasing by 13\% compared to their parent seeds on average. 
This inefficiency underscores the need for a dedicated mutation strategy tailored to the target code, which we believe could significantly enhance the efficiency of directed fuzzing.
\end{itemize}

\section{Background}

In this section, we provide background information about directed grey-box fuzzing, with a focus on the calculation method and granularity of distance metrics.

\subsection{Directed Grey-box Fuzzing}

\begin{algorithm}
    \floatname{algorithm}{Algortihm}
    \renewcommand{\algorithmicrequire}{\textbf{Input:}}
    \renewcommand{\algorithmicensure}{\textbf{Output:}}
    \caption{Directed Grey-box Fuzzing}
    \label{alg1}
    \begin{algorithmic}[1]
        \Require Initial Seed Pool $S$
        \Require Target Code $T$
        \Require Program $P$
        \Ensure Crash $C$
        \While {not $Timeout$}
            \State {$s$ = $ChooseNext(S)$ }
            \State {$d$ = $Distance(s, T)$ }
            \State {$p$ = $AssignEnergy(s, d)$}
            \For{$1$ $\leq$ $t$ $\leq$ $p$}
                \State{$testcase$ $=$ $Mutate(s)$}
                \State{$Run(P, testcase)$}
                \If{$P$ $crashes$}
                    \State{ add $testcase$ to $C$}
                \EndIf
                \If{ $Interesting(testcase)$}
                    \State{ add $testcase$ to $S$}
                \EndIf
            \EndFor
        \EndWhile
    \end{algorithmic}
\end{algorithm}

Directed grey-box fuzzing is an iterative process that approaches the target code by prioritizing seeds with shorter distances. 
Initially, as depicted in Algorithm \ref{alg1}, the initial seed pool $S$ and the location of the target code $T$ are provided by the user. As the fuzzing loop progresses, seed $s$ is selected from the seed pool $S$, and its distance $d$ to the target code is calculated. 
The number of testcases generated from the selected seed $s$ is determined by its distance $d$. 
Seeds with shorter distances receive more energy and more testcases are generated from them. As a result, the probability of newly generated testcases approaching the target code increases until the target vulnerability is triggered.

For instance, AFLGo calculates seed distance based on the call graph and control flow graph, and then assigs energy to seeds according to their distance and the default power schedule strategy of AFL. 
This design, which places significant emphasis on distance metric, is echoed in related works such as WindRanger~\cite{Windranger}, Beacon~\cite{beacon}, DAFL~\cite{DAFL}, and SelectFuzz~\cite{SelectFuzz}.

\subsection{Distance Calculation}\label{sec:distance_calculation}

The distance of a seed is the distance between the basic blocks in the execution trace and the basic blocks where the target code resides. 
Generally, it is calculated by compounding function-level distance and basic-block-level distance among those basic blocks.
Here we take AFLGo as an example to illustrate the details of calculating distance.

\subsubsection{Function-Level Distance}

The function-level distance is the distance between a given function to all the target functions (the target code may reside in many functions) in the call graph. 
Formally, the function distance $d_f(f_1,f_2)$ between any two functions $f_1$ and $f_2$ is defined as the number of edges along the shortest path from $f_1$ to $f_2$ in the call graph.
In AFLGo, the function distance between a given function $f$ and target functions $T_f$ is defined as the harmonic mean of the function distance between $f$ and every target function $t_f \in T_f$:
\begin{equation}\label{eq:background_func_dis}
d_f(f,T_f)= \begin{cases}
\textit{undefined} & R(f,T_f)=\emptyset \\
\left[{\sum}_{t_f \in R(f,T_f)} d_f(f,t_f)^{-1}\right]^{-1} & R(f,T_f)\neq\emptyset 
\end{cases}
\end{equation}
where $R(f,T_f)$ is the set of target functions that are reachable from $f$ in call graph.

\subsubsection{Basic-Block-Level Distance}\label{sec:BBLD}

The basic-block-level distance $d_b(b, T_b)$ is the distance between a given basic block $b$ and the target basic blocks $T_b$ (the target code may reside in many basic blocks).

Specifically, if the given basic block $b$ is one of the target basic blocks, $d_b(b, T_b)$ is set to zero. If the basic block $b$ calls a function $f$ via which could reach target functions, $d_b(b, T_b)$ is set to a multiple of the function-level distance of $f$. Moreover, if the basic block $b$ calls more than one function, the function with the shortest function-level distance to the target functions will be chosen. These functions, which are called by basic blocks and in the call chain towards target functions, are denoted as $R_f$. As for basic blocks that call the function $f \in R_f$, they are denoted as $R_b$.

Otherwise, even if the basic block $b$ doesn't call any function in $R_f$, it may reach target basic blocks via another basic block $r \in R_b$. Therefore, $d_b(b, T_b)$ is defined as the sum of $d_b(b, r)$, and $d_b(r, T_b)$. 

Formally, the basic-block-level distance $d_b(b, T_b)$ between a basic block $b$ and the target basic blocks $T_b$ is defined as:
\begin{equation}\label{eq:background_bb_dis}
d_b(b,T_b) = \begin{cases}
0 &  b \in T_b \\
c \times \min_{f \in R_f}(d_f(f,T_f)) & b \in R_b \\ 
\left[ \sum_{r \in R_b} (d_b(b,r) + d_b(r,T_b))^{-1} \right]^{-1} & \mathrm{otherwise}
\end{cases}
\end{equation}
where $c$ is a constant to magnify function-level distance in AFLGo.

\subsubsection{Seed Distance}

The execution trace of a seed $s$ consists of many basic blocks exercised in one fuzzing round. Let $\phi(s)$ be the set of basic blocks in the seed trace of $s$. The seed distance $d_s(s, T_b)$ is defined as:
\begin{equation}\label{eq:background_arithmetic_trace_dis}
d_s(s, T_b) = \frac{{\sum}_{b \in \phi(s)}d_b(b, T_b)}{|\phi(s)|}
\end{equation}
where $|\phi(s)|$ is the basic block number in the execution trace of seed $s$.

\section{Distance Calculation Method and Granularity}\label{sec:defi_and_granu}\label{sec:method_and_granularity}

Although the aforementioned calculating procedure of distance metrics is widely acknowledged, there are many modifications and improvements introduced by other works \cite{1dVul, CAFL, hawkeye, SelectFuzz, TargetFuzz, Windranger, G-Fuzz}.
Here we discuss the details of typical distance metrics from the perspective of distance calculation method and distance granularity.

\subsection{Distance Calculation Method}

\begin{figure*}[htbp]
    \centering
    \subfloat[Arithmetic Mean Distance]{
    \includegraphics[width=0.2\linewidth]{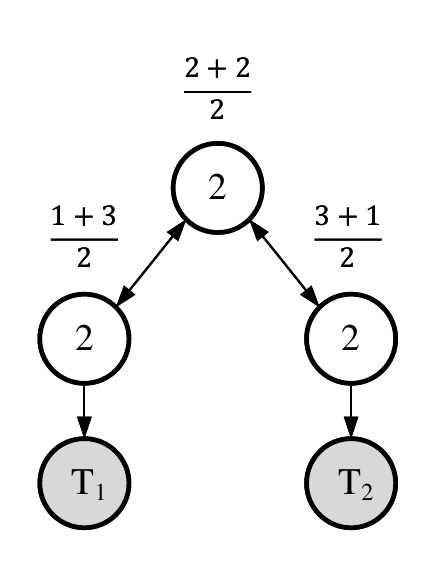}
    }
    \quad
    \subfloat[Harmonic Mean Distance]{
    \includegraphics[width=0.205\linewidth]{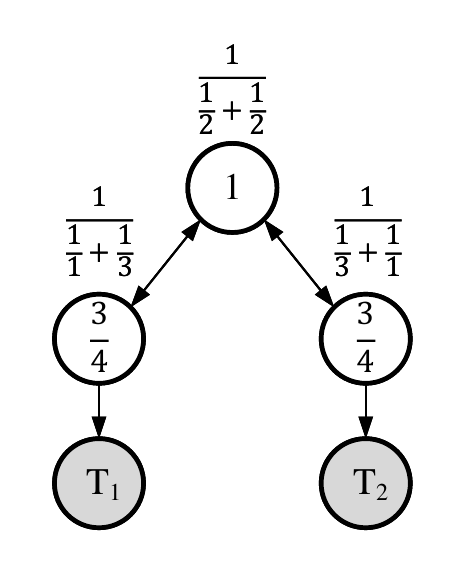}
    }
    \quad
    \subfloat[Closest Target Distance]{
    \includegraphics[width=0.215\linewidth]{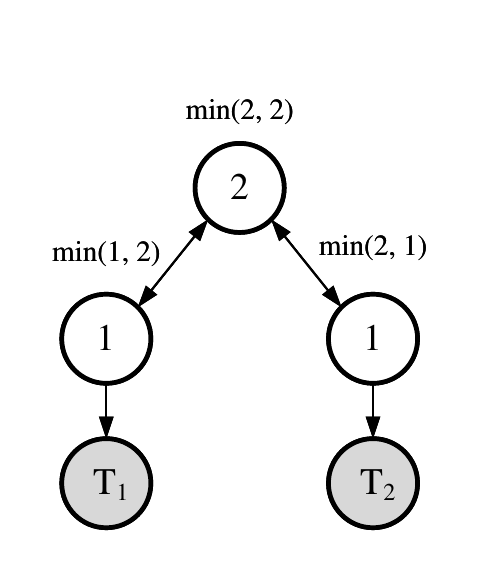}
    }
    \caption{Difference among arithmetic mean distance, harmonic mean distance, and closest target distance. Node distance is shown in the white circles. The target nodes are marked in gray.}
    \label{fig:arithmetic_and_harmonic_example}
\end{figure*}

The seed distance is calculated by compounding code distance (function-level distance and basic-block-level distance). 
Overall, there are three different distance calculation methods to obtain these distance metrics.

\subsubsection{Arithmetic Mean Distance}

Intuitively, the distance between one and multiple nodes could be defined as the arithmetic mean of the distance between the given node and each target node.
Correspondingly, the function-level distance $d_f(f,T_f)$ is defined as the arithmetic mean of the distance between the given function and each target function:
\begin{equation}\label{eq:defi_arithmetic_func_dis}
d_f(f,T_f)= \begin{cases}
\textit{undefined} & R(f,T_f)=\emptyset \\
\frac{{\sum}_{t_f \in R(f,T_f)} d_f(f,t_f)}{|R(f,T_f)|} & R(f,T_f)\neq\emptyset 
\end{cases}
\end{equation}
where $T_f$ is the set of target functions and $R(f,T_f)$ is the set of target functions that are reachable from f in the call graph.

The basic-block-level distance $d_b(b, T_b)$ is defined as:
\begin{equation}\label{eq:defi_arithmetic_bb_dis}
d_b(b,T_b) = \begin{cases}
0 &  b \in T_b \\
c \times \min_{f \in R_f}(d_f(f,T_f)) & b \in R_b \\ 
\frac{\sum_{r \in R_b} (d_b(b,r) + d_b(r,T_b))}{|R_b|} & \mathrm{otherwise}
\end{cases}
\end{equation}
Note that the $R_f$ refers to functions called by basic block $b$ and in the call chain towards target functions. 
The $R_b$ refers to basic blocks that call function $f \in R_f$.

\subsubsection{Harmonic Mean Distance}

The arithmetic mean distance is simple and intuitive but cannot distinguish a node that is closer to one target and further from another from a node that is equidistant from both targets. As shown in Fig.~\ref{fig:arithmetic_and_harmonic_example}, the same distance will be assigned to each node if the arithmetic mean distance is adopted.

Therefore, AFLGo adopts the harmonic mean distance to assign a shorter distance to a node that is closer to one target. The corresponding equations to define the function-level distance and basic-block-level distance are presented as (\ref{eq:background_func_dis}) and (\ref{eq:background_bb_dis}).

\subsubsection{Closest Target Distance}

Assigning the path length to the closest target as the distance of a basic block is another way to define the distance of a basic block or function to targets. Only the closest target basic block or function counts while other further ones are ignored in this distance calculation method. Correspondingly, the function-level distance $d_f(f,T_f)$ is defined as:
\begin{equation}\label{eq:defi_func_shortest_path}
d_f(f,T_f)= \begin{cases}
\textit{undefined} &  R(f,T_f)=\emptyset \\
\min_{t_f \in R(f,T_f)}(d_f(f,t_f))   &  R(f,T_f)\neq\emptyset 
\end{cases} 
\end{equation}
and the basic-block-level distance $d_b(b, T_b)$ is defined as:
\begin{equation}\label{eq:defi_bb_shortest_path}
d_b(b,T_b) = \begin{cases}
0 &  b \in T_b \\
c \times \min_{f \in R_f}(d_f(f,T_f)) & b \in R_b \\ 
\min_{r \in R_b} (d_b(b,r) + d_b(r,T_b)) & \mathrm{otherwise}
\end{cases}
\end{equation}

Moreover, those distance calculation methods are implemented in the stage of calculating the seed distance.
Specifically, the arithmetic mean of the distance of each basic block in the execution trace is the distance of the seed in the arithmetic mean distance calculation method.
The harmonic mean of the distance of each basic block in the execution trace is the distance of the seed in the harmonic mean distance calculation method.
The smallest distance of all the basic blocks in the execution trace is the distance of the seed in the closest target distance calculation method.

\subsection{Distance Calculation Granularity}

Aside from the various distance calculation methods, different granularities may lead to different effectiveness of distance metrics in guiding directed fuzzing.
For example, the approximated basic-block-level distance based on compounding function-level distance and basic-block-level distance in AFLGo may outperform that based on mere function-level distance.
Here we discuss three different distance calculation granularities: function-level distance, approximated distance, and fine-grained distance.

\subsubsection{Approximated Basic-Block-Level Distance}

Most of the existing works approximate basic-block-level distance by simply multiplying function-level distance with a constant as (\ref{eq:background_bb_dis}).

However, the constant $c$ ($c$=10 in AFLGo) may introduce bias, affecting the accuracy of the result. On one hand, a fixed constant cannot fit the fact that the number of basic blocks may vary greatly in different functions. On the other hand, when there are many functions along the call chain from the given basic block to the target code, the function-level distance dominates the overall distance while the basic-block-level distance within the function is negligible.

\subsubsection{Function-Level Distance}

Given the domination of function-level distance in the calculation, distance semantics mainly remain by replacing the basic-block-level distance with function-level distance. Formally, the distance $d_b(b, T_b)$ between the given basic block $b$ and the target basic blocks $T_b$ is defined as:
\begin{equation}\label{eq:granu_func_dis}
d_b(b,T_b) = d_f(f, T_f)
\end{equation}
where $f$ is the function where the given basic block $b$ resides and $T_f$ is the target function set.

\subsubsection{Basic-Block-Level Distance}

To address the problem of the fixed constant $c$, algorithms that obtain the fine-grained distance are proposed by researchers. Generally, these algorithms focus on calculating more precise distances between adjacent functions by analyzing the call pattern \cite{hawkeye, CAFL, Windranger}.

In this paper, we propose an algorithm to calculate the fine-grained distance between adjacent functions by analyzing all potential call paths. First, for caller function $A$ and callee function $B$, all the basic blocks that call function $B$ within the control flow graph of function $A$ will be recorded. Second, the distance $d_b(e,b)$ between the entrance basic block $e$ and the basic block $b$ of the function $A$ is defined as the number of edges along the shortest path from $e$ to $b$ in the control flow graph of the function $A$. Third, the function distance $d_f(A, B)$ between function $A$ and function $B$ is defined as the shortest distance of paths between entrance basic block $e$ and every basic block that calls function $B$:
\begin{equation}\label{eq:granu_fine_bblk}
d_f(A,B)= min_{b \in CallSite(A,B)}d_b(e,b)
\end{equation}
where $CallSite(A,B)$ is the set of basic blocks that call function $B$ in the control flow graph of function $A$.

Consequently, the weight of each edge in the call graph is the distance between corresponding adjacent functions. The function distance $d_f(f_1, f_2)$ between any two functions $f_1$ and $f_2$ is defined as the sum of the weight of edges along the shortest path between $f_1$ and $f_2$.

Moreover, the distance $d_b(b, T_b)$ is defined as:
\begin{equation}\label{eq:granu_fine_bblk_to_bb_dis}
d_b(b,T_b) = \begin{cases}
0 &  b \in T_b \\
\min_{f \in R_f}(d_f(f,T_f)) & b \in R_b\\ 
\left[ \sum_{r \in R_b} (d_b(b,r) + d_b(r,T_b))^{-1} \right]^{-1} & \mathrm{otherwise}
\end{cases} 
\end{equation}

\section{Evaluation}

We implement the aforementioned different distance metrics based on AFLGo. 
Then, we conduct comprehensive experiments to evaluate the performance of these distance metrics and answer the research questions raised in section~\ref{sec:intro}:
\begin{itemize}
    \item \textbf{RQ1}: How do different distance calculation methods perform in practice?
    \item \textbf{RQ2}: Does the distance with finer granularity perform better in guiding directed fuzzing?
    \item \textbf{RQ3}: Do the seeds with closer distance have more contribution in generating PoC?
\end{itemize}
Moreover, we quantitively investigate the capability of the current mutation strategy of DGF in generating better testcases.

\subsection{Implementation}

For the distance calculation method, we implement \textit{arithmetic mean distance}, \textit{harmonic mean distance}, and \textit{closest target distance} based on AFLGo. For the distance granularity, we implement the \textit{approximated distance}, \textit{function-level distance}, and \textit{basic-block-level distance}. 
As for \textit{approximated distance}, we use $c=10$ to magnify the function-level distance.

For convenience, we use the following abbreviations to represent different distance calculation methods and granularities: \textbf{arithmetic} represents \textit{arithmetic mean distance}, \textbf{harmonic} represents \textit{harmonic mean distance}, \textbf{closest} represents \textit{closest target distance}, \textbf{appr} represents \textit{approximated basic-block-level distance}, \textbf{func} represents \textit{function-level distance}, and \textbf{bblk} represents \textit{basic-block-level distance}.

\subsection{Experiment Settings}

\textbf{Experiemnt Environment.}
We conducted all the experiments on the server with Ubuntu 20.04 equipped with 26 Intel(R) Xeon(R) Gold 6230R CPU cores with 2.10 GHz and 220 GB of memory. 
Considering the randomness inherent in fuzzing, each evaluation is repeated 10 times. 

\textbf{Benchmark.}
We choose the reproducible vulnerabilities to mitigate potential experimental basis, as shown in Table~\ref{tab:benchmark}.
These vulnerabilities are used in previous works, such as AFLGo, Hawkeye, Windranger, and DAFL.
Binutils~\cite{Binutils24:online} is a collection of binary analysis tools. 
libxml2~\cite{GNOMElib85:online} is a software library for parsing XML documents. 
libming is a C library for creating Adobe Flash (.swf) files from PHP, C++, and many other languages. 
Binutils and libxml2 have almost one million lines of code. 
All of them are widely used open-source projects. 
All the vulnerabilities are identified by the CVE-ID and well documented in the US National Vulnerability Data. 
In experiments, the time-to-exploitation is set to 21 hours and the timeout is set to 24 hours. 

\begin{table}[t]
\caption{Benchmark Composition.}
\label{tab:benchmark}
\begin{center}
\begin{threeparttable}
\begin{tabular}{cccc}
\hline
\textbf{Package}              & \textbf{Program}             & \textbf{CVE-ID} & \textbf{Previous Research} \\
\hline
\multirow{4}{*}{Binutils}     &   \multirow{4}{*}{cxxfilt}   &   2016-4487   &   \multirow{4}{*}{\makecell{AFLGo, Beacon, WindRanger\\Hawkeye, ParmeSan, DAFL\\UAFuzz}}      \\ 
                              &                              &   2016-4489   &     \\ 
                              &                              &   2016-4490   &     \\ 
                              &                              &   2016-4492   &   \\
\hline
\multirow{2}{*}{libming}      &   \multirow{2}{*}{swftophp}  &  \multirow{2}{*}{2018-8807}   &   \multirow{2}{*}{\makecell{AFLGo, Beacon, WindRanger\\DAFL, SelectFuzz}}   \\
                              &                              &           &      \\
\hline
\multirow{2}{*}{libxml2}      &   \multirow{2}{*}{xmllint}   &   \multirow{2}{*}{2017-9047}   &   \multirow{2}{*}{\makecell{AFLGo, Beacon\\ DAFL, SelectFuzz}}   \\
                              &                              &              & \\
\hline
\end{tabular}
\end{threeparttable}
\end{center}
\end{table}

\textbf{Evaluation Metrics.}\label{sec:evalmetric}
We determine whether a testcase exposes the target vulnerability by executing the failing inputs on the patched version of the target program. 
If a failing testcase passes on the patched version, it is said to witness the target vulnerability. 

We use the following measures of fuzzing efficiency and performance gain. 
\textit{Time-To-Exposure} (TTE) measures the time consumption of the fuzzing campaign until the first testcase is generated that exposes the target vulnerability. 
For a run that doesn't expose the target vulnerability, the timeout (24h) will be recorded as the TTE. 

The \textit{factor improvement} (Factor) measures the performance gain as the mean TTE of algorithm B divided by that of algorithm A. 
It indicates that A outperforms B if the Factor is greater than one. 

The \textit{Vargha-Delaney statistic} ($\hat{A}_{12}$)~\cite{Vargha2000ACA} is a non-parametric measure of effect size and is also the recommended standard measure for the evaluation of randomized algorithms. 
Given a performance measure M (such as TTE) seen in m measures of algorithm A and n measures of algorithm B, the $\hat{A}_{12}$ statistic measures the probability that using algorithm A yields higher M values than using algorithm B. We use \textit{Mann-Whitney U}~\cite{mann-whitney-u} to measure the statistical significance of performance gain. When it is significant ($p-value \le 0.05$), we mark the corresponding $\hat{A}_{12}$ values in bold.

\subsection{Performance Analysis on Different Distance Calculation Methods and Granularities}

In this experiment, we implement 9 combinations of distance calculation methods and granularities on the AFLGo. 
The performance of different distance calculation methods and granularities is demonstrated by comparing the mean TTE ($\mu$TTE) to expose the target vulnerability of different algorithms.

\begin{table}[t]
\small
\caption{Performance Comparison among Different Distance Calculation Methods.}
\label{tab:methods}
\begin{center}
\begin{threeparttable}
\begin{tabular}{ccccccc}
\hline
    CVE-ID                      & Method     & Runs  & $\mu$TTE        & Factor    & $\hat{A}_{12}$ \\ \hline
    \multirow{3}{*}{2016-4487}  & harmonic   & 10    & 291             & -         & - \\
                                & arithmetic & 10    & \textbf{114.26} & 2.55      & 0.56 \\ 
                                & closest    & 10    & 142.52          & 2.04      & 0.56 \\ \hline
    \multirow{3}{*}{2016-4489}  & harmonic   & 9     & 258.51          & -         & - \\
                                & arithmetic & 10    & \textbf{139.72} & 1.85      & 0.75 \\ 
                                & closest    & 10    & 146.59          & 1.76      & 0.54 \\ \hline
    \multirow{3}{*}{2016-4490}  & harmonic   & 10    & \textbf{32.62}  & -         & - \\ 
                                & arithmetic & 10    & 34.26           & 0.95      & 0.49 \\ 
                                & closest    & 10    & 44.47           & 0.73      & 0.30 \\ \hline
    \multirow{3}{*}{2016-4492}  & harmonic   & 9     & 451.62          & -         & - \\ 
                                & arithmetic & 8     & \textbf{390.66} & 1.16      & 0.64 \\ 
                                & closest    & 9     & 489.39          & 0.92      & 0.44 \\ \hline
    \multirow{3}{*}{2018-8807}  & harmonic   & 9     & 1002.95         & -         & - \\ 
                                & arithmetic & 9     & \textbf{869.02} & 1.15      & 0.64 \\ 
                                & closest    & 8     & 1069.13         & 0.94      & 0.49 \\ \hline
    \multirow{3}{*}{2017-9047}  & harmonic   & 10    & 223.08          & -         & - \\ 
                                & arithmetic & 10    & 162.33          & 1.37      & 0.58 \\ 
                                & closest    & 10    & \textbf{97.45}  & 2.29      & 0.68 \\ \hline
    \multirow{3}{*}{}           &      & \multicolumn{2}{r}{Mean Factor}& \multicolumn{2}{r}{Mean $\hat{A}_{12}$}  \\
                                & arithmetic  & \multicolumn{2}{r}{1.51}& \multicolumn{2}{r}{0.61} \\
                                & closest    & \multicolumn{2}{r}{1.45}& \multicolumn{2}{r}{0.50} \\
\hline
                                
\end{tabular}
\begin{tablenotes}
\footnotesize
\item[1] The distance graunlarity is set to approximated basic-block-level (appr) distance.
\item[2] The statistically significant values of $\hat{A}_{12}$ and the smallest TTE is highlighted in bold. 
\item[3] A run that does not reproduce the vulnerability within 24 hours receives a TTE of 24 hours.
\end{tablenotes}
\end{threeparttable}
\end{center}
\end{table}

\begin{table}[t]
\small
\caption{Performance Comparison among Different Distance Granularities.}
\label{tab:granularities}
\begin{center}
\begin{threeparttable}
\begin{tabular}{ccccccc}
\hline
CVE-ID                      & Granularity   & Runs  & $\mu$TTE          & Factor    & $\hat{A}_{12}$ \\ \hline
\multirow{3}{*}{2016-4487}  & func          & 10    & 126.81            & -         & - \\ 
                            & appr          & 10    & \textbf{114.26}   & 1.11      & 0.51 \\ 
                            & bblk          & 10    & 148.02            & 0.86      & 0.61 \\ \hline
\multirow{3}{*}{2016-4489}  & func          & 10    & 219.64            & -         & - \\ 
                            & appr          & 10    & \textbf{139.72}   & 1.57      & 0.81 \\
                            & bblk          & 9     & 241.59            & 0.91      & 0.56 \\ \hline
\multirow{3}{*}{2016-4490}  & func          & 10    & \textbf{32.15}    & -         & - \\ 
                            & appr          & 10    & 34.26             & 0.94      & 0.45 \\ 
                            & bblk          & 10    & 35.36             & 0.91      & 0.43 \\ \hline
\multirow{3}{*}{2016-4492}  & func          & 9     & 376.36            & -         & - \\
                            & appr          & 8     & 390.66            & 0.96      & 0.66 \\ 
                            & bblk          & 10    & \textbf{268.32}   & 1.40      & 0.53 \\ \hline
\multirow{3}{*}{2018-8807}  & func          & 10    & \textbf{837.95}   & -         & - \\ 
                            & appr          & 9     & 869.02            & 0.96      & 0.50 \\ 
                            & bblk          & 9     & 953.46            & 0.88      & 0.40 \\ \hline
\multirow{3}{*}{2017-9047}  & func          & 10    & 111.19            & -         & - \\ 
                            & appr          & 10    & 162.33            & 0.68      & 0.44 \\ 
                            & bblk          & 10    & \textbf{80.21}    & 1.39      & 0.56 \\ \hline
\multirow{3}{*}{}           &          & \multicolumn{2}{r}{Mean Factor}& \multicolumn{2}{r}{Mean $\hat{A}_{12}$}  \\
                            & appr            & \multicolumn{2}{r}{1.04}& \multicolumn{2}{r}{0.56} \\
                            & bblk            & \multicolumn{2}{r}{1.06}& \multicolumn{2}{r}{0.52} \\
\hline

\end{tabular}
\begin{tablenotes}
\footnotesize
\item[1] The distance calculation method is set to the arithmetic mean.
\item[2] The statistically significant values of $\hat{A}_{12}$ and the smallest TTE is highlighted in bold. 
\item[3] A run that does not reproduce the vulnerability within 24 hours receives a TTE of 24 hours.
\item[4] func, appr, bblk is short for function-level distance, approximated basic-block-level distance, and basic-block-level distance respectively.
\end{tablenotes}
\end{threeparttable}
\end{center}
\end{table}

As shown in Table~\ref{tab:methods}, fuzzing campaigns that use \textit{arithmetic mean distance} to calculate distance expose 4 vulnerabilities with the fastest speed while those that use \textit{harmonic mean distance} expose 3 vulnerabilities with the slowest speed. 
Overall, the distance based on \textit{arithmetic mean distance} surprisingly outperforms that based on \textit{harmonic mean distance} 1.51 times on average, and the distance based on \textit{closest target distance} outperforms that based on \textit{harmonic mean distance} 1.45 times on average.
The merit of the \textit{harmonic mean distance} doesn't promote the performance of fuzzing as the design philosophy of AFLGo.

However, considering that $\hat{A}_{12}$ in most cases fail into $(0.36, 0.64)$, the probability of \textit{arithmetic mean distance} and \textit{closest target distance} outperform \textit{harmonic mean distance} is small even negligible.
Moreover, there is no statistically significant value of $\hat{A}_{12}$ with $p-value < 0.05$, thus the advantage of \textit{arithmetic mean distance} and \textit{closest target distance} is difficult to sustain when their reproductivity is measured by the \textit{Vargha Delaney tatistic}. 

Therefore, it still lacks solid evidence to prove that the \textit{arithmetic mean distance} or the \textit{closest target distance} is more suitable to calculate distance than the \textit{harmonic mean distance}.

As shown in Table~\ref{tab:granularities}, \textit{func}, \textit{appr} and \textit{bblk} guides fuzzing to expose 2 target vulnerabilities with the fastest speeds respectively though they calculate the distance in different granularity. 
According to the definition in \ref{sec:evalmetric}, the Mean Factor of \textit{bblk} and \textit{appr} over \textit{func} is quite close to 1, indicating that there is no significant performance gain by enhancing the distance granularity.
Moreover, there is no statistically significant performance gain since all of the p-values are greater than 0.05.
Therefore, there is no global trend about which distance granularity is the best one to guide fuzzing across these CVEs.

Although a specific distance calculation method or granularity may yield the shortest TTE in some cases, the performance gain remains trivial when measuring the reproductivity of such a performance by the \textit{Vargha-Delaney statistic} and the \textit{Mann-Whitney U} test. 
Therefore, there is no steady trend about which distance calculation method or granularity is the best one to guide the fuzzing. 

In other words, the effect of distance metrics in directed fuzzing has encountered a bottleneck. 
While various distance calculation methods and granularities have been explored, the improvements in fuzzing performance are always marginal. 
It appears that the potential gains from refining distance calculation method or granularity are limited. 
To achieve substantial advancements, the most promising approach is to incorporate additional information beyond mere distance metrics.

\begin{tcolorbox}[arc=2mm,title = {Lesson 1}]

For RQ1 and RQ2, the difference among different distance metrics is not very significant.
Therefore, it is necessary to design more effective metrics to guide the directed grey-box fuzzing.

\end{tcolorbox}

\subsection{Lineages Analysis of Proof-of-Concept}

\begin{figure*}[!thbp]
\centering
\captionsetup[subfloat]{labelformat=empty}
\subfloat[CVE-2016-4487]{\includegraphics[width=0.16\linewidth]{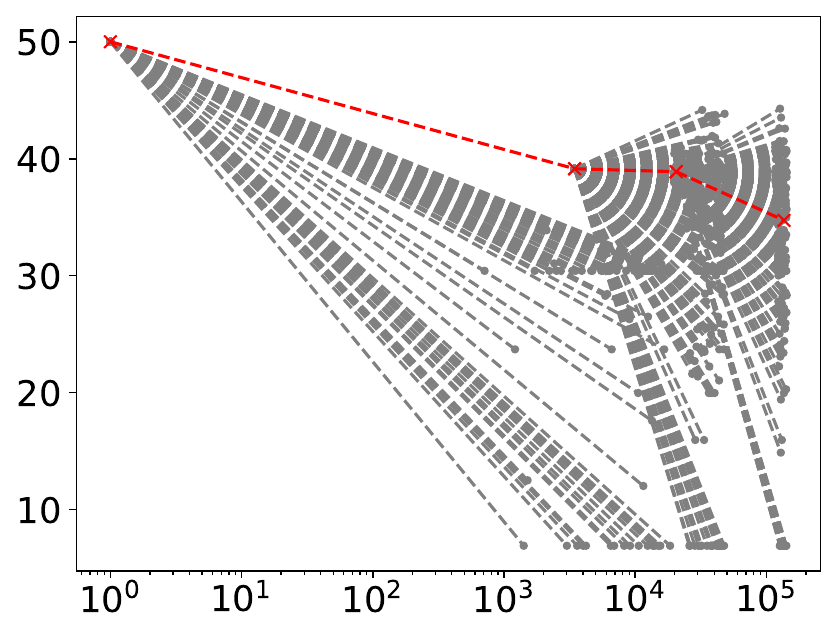}}
\subfloat[CVE-2016-4489]{\includegraphics[width=0.16\linewidth]{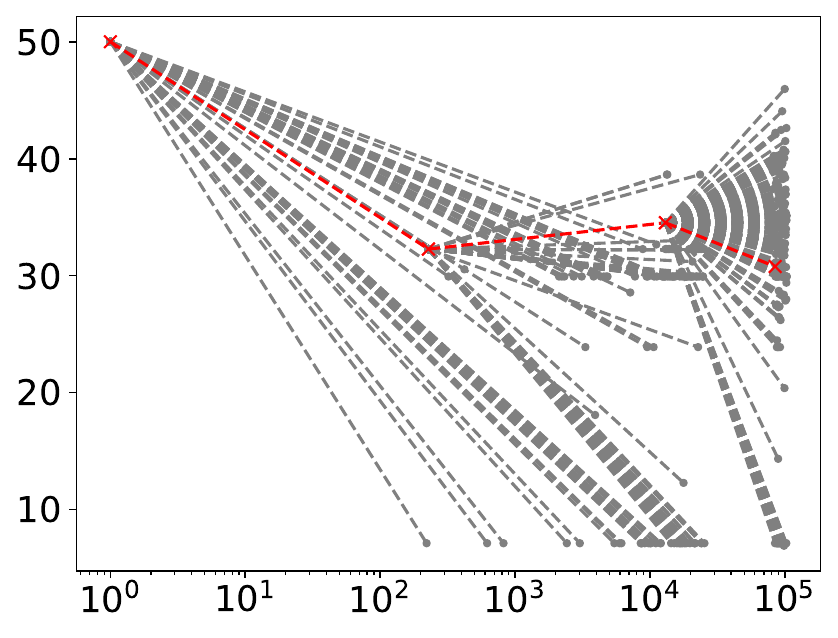}}
\subfloat[CVE-2016-4490]{\includegraphics[width=0.16\linewidth]{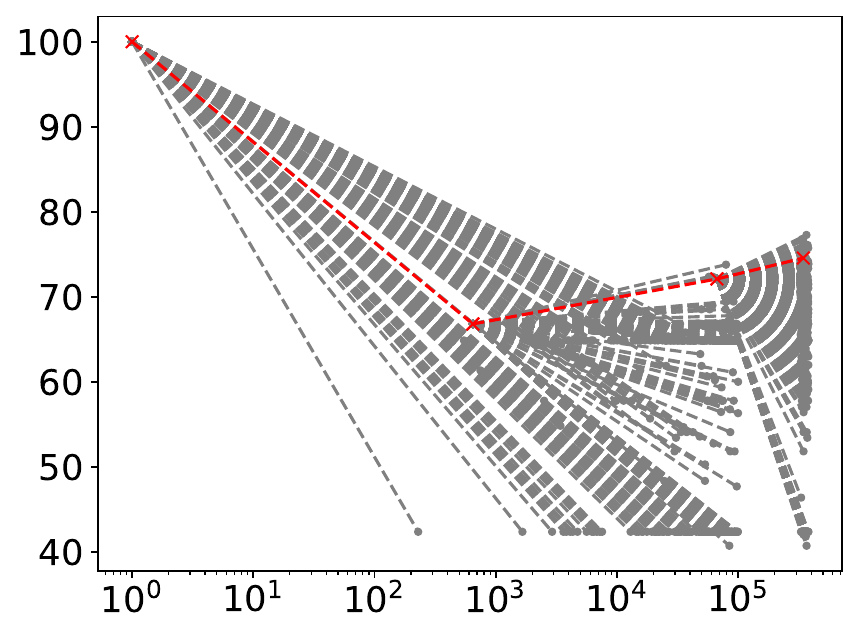}}
\subfloat[CVE-2016-4492]{\includegraphics[width=0.16\linewidth]{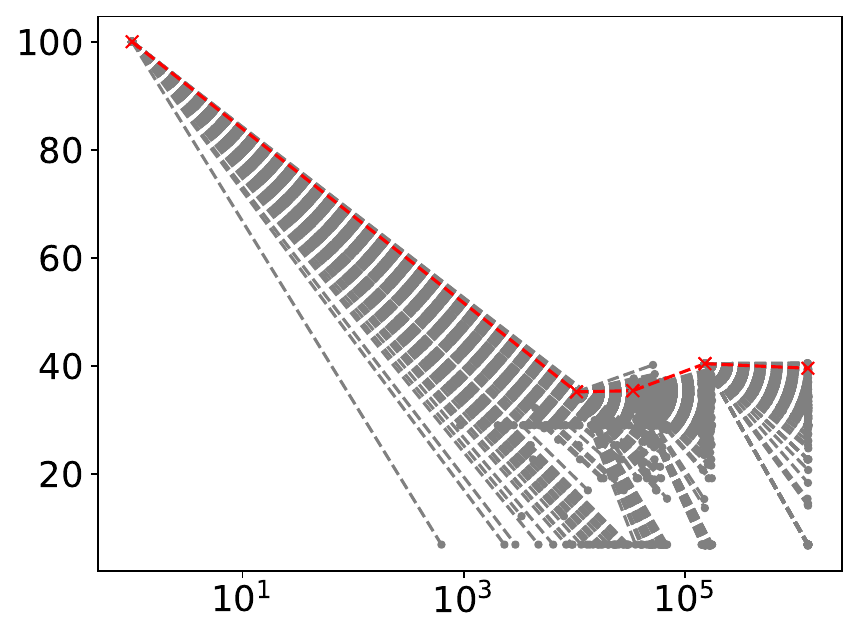}}
\subfloat[CVE-2018-8807]{\includegraphics[width=0.153\linewidth]{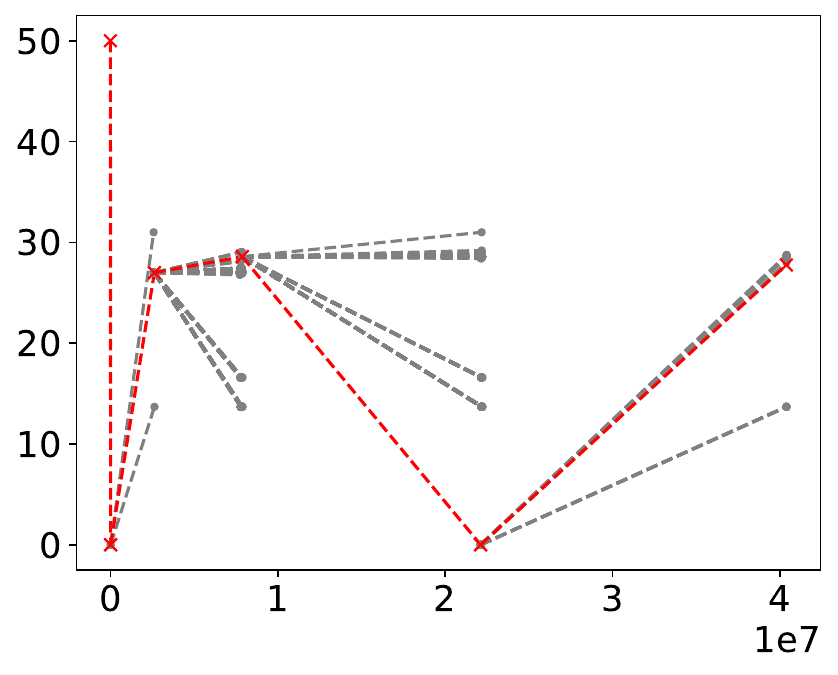}}
\subfloat[CVE-2017-9047]{\includegraphics[width=0.156\linewidth]{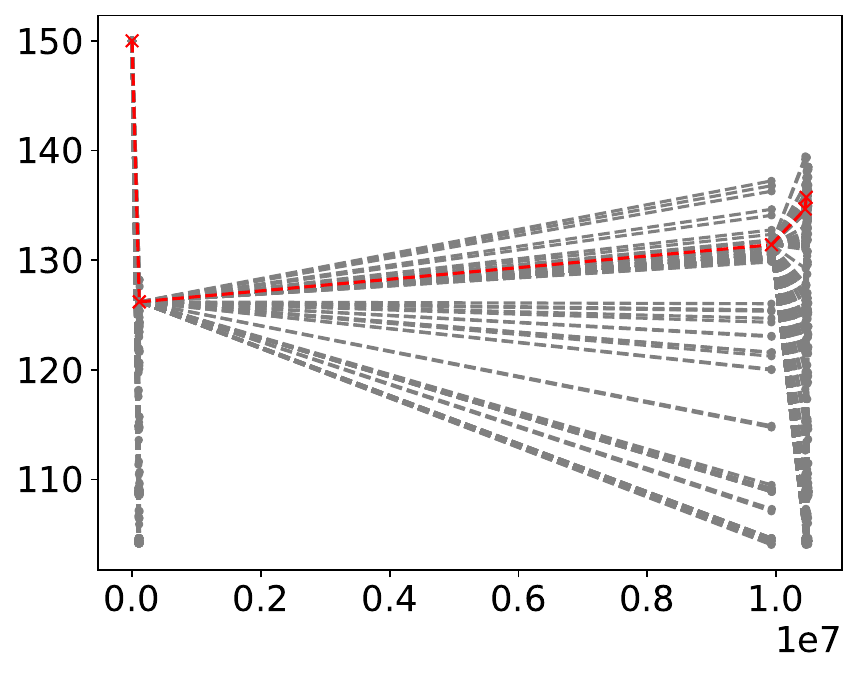}}
\caption{Lineage of PoC on different CVEs in the benchmark. Red 'x' nodes represent all the ancestor seeds of the PoC, with the rightmost red 'x' corresponding to the PoC itself. Grey nodes represent testcases derived from ancestor seeds of PoC. The x-axis represents the time (ms) when the seed or testcase is generated and the y-axis represents their distance.}
\label{fig:lineage_of_PoC}
\end{figure*}

A comprehensive analysis of the lineages of PoC (Proof-of-Concept) is conducted to gain insight into the evolution from the initial seeds to the PoC in the iterative fuzzing loop guided by distance metric. 

\subsubsection{Short Lineage Length}

\begin{figure}[!t]
\centerline{\includegraphics[width=0.9\linewidth]{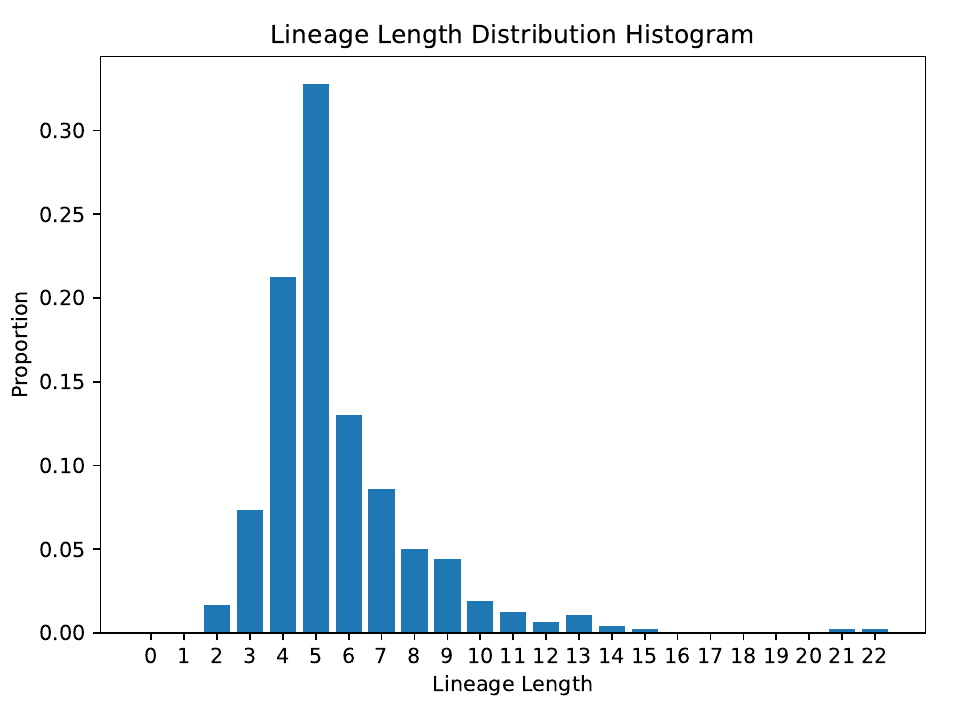}}
\caption{The Distribution of Lineage Length of PoC.}
\label{fig:lineage_length_hist}
\end{figure}

Many testcases will be yielded and reported as input which crashes the target program in one fuzz run. 
If a failing testcase passes on the patched version of the target program, it is said to witness the target vulnerability. 
Firstly, the PoC that exposes the target vulnerability is identified. 
If multiple  PoCs are identified, only the first one is retained. Secondly, we back trace all the ancestors of PoC by analyzing the name of each seed and crash. 
AFLGo records all the seeds and crashes in a naming format that could indicate the origin of the seed from which the current seed or crash is mutated. 
The initial seed is the ancestor of every seed and crash and is manually designated as empty or identical among every fuzz run. 
Thirdly, for every PoC uncovered in each fuzz run across all CVEs in our dataset, the number of ancestor seeds (lineage length) is recorded and the lineage length distribution histogram is illustrated as Fig.~\ref{fig:lineage_length_hist}.

The majority of PoCs originate from lineages with a length of 7 or fewer generations.
On the one hand, it underscores that only a few generations of mutation are required to derive a PoC from the initial seed. 
On the other hand, the pivotal role of mutation strategy is highlighted in triggering the target vulnerability in directed fuzzing.

\subsubsection{Deviation between Distance and Reachability}\label{sec:poc_lineage_analysis}

\begin{figure*}[!t]
\centering
\captionsetup[subfloat]{labelformat=empty}
\subfloat[CVE-2016-4487]{\includegraphics[width=0.177\linewidth]{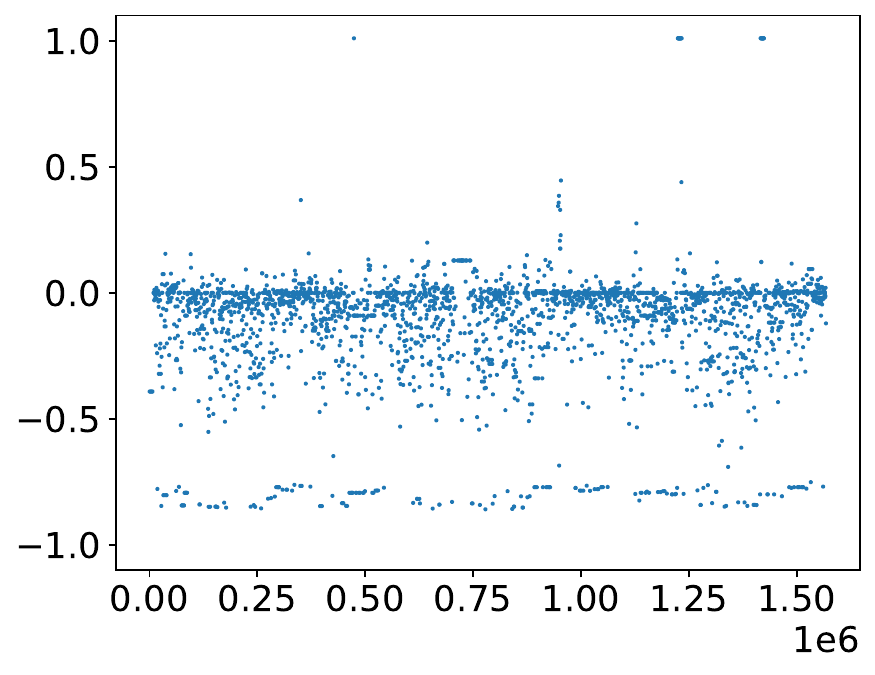}}
\subfloat[CVE-2016-4489]{\includegraphics[width=0.16\linewidth]{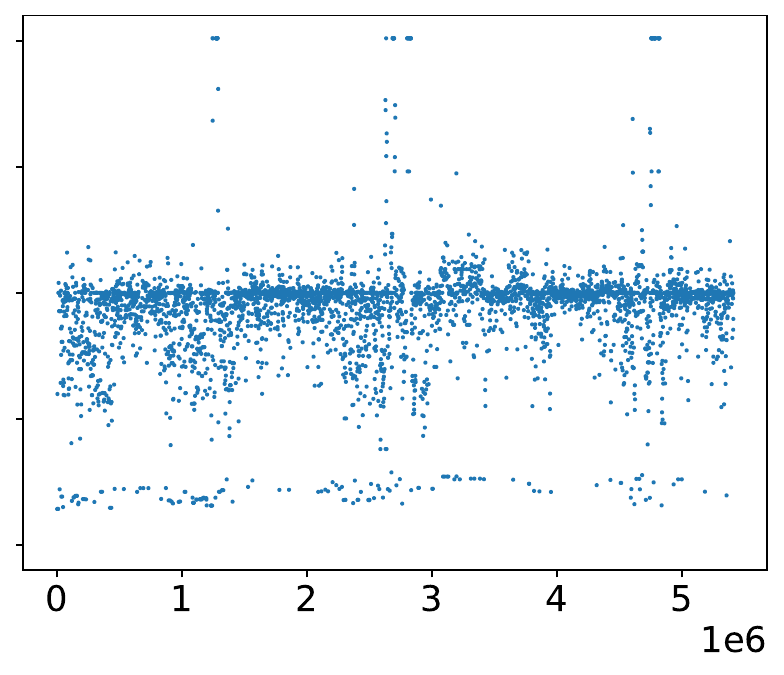}}
\subfloat[CVE-2016-4490]{\includegraphics[width=0.16\linewidth]{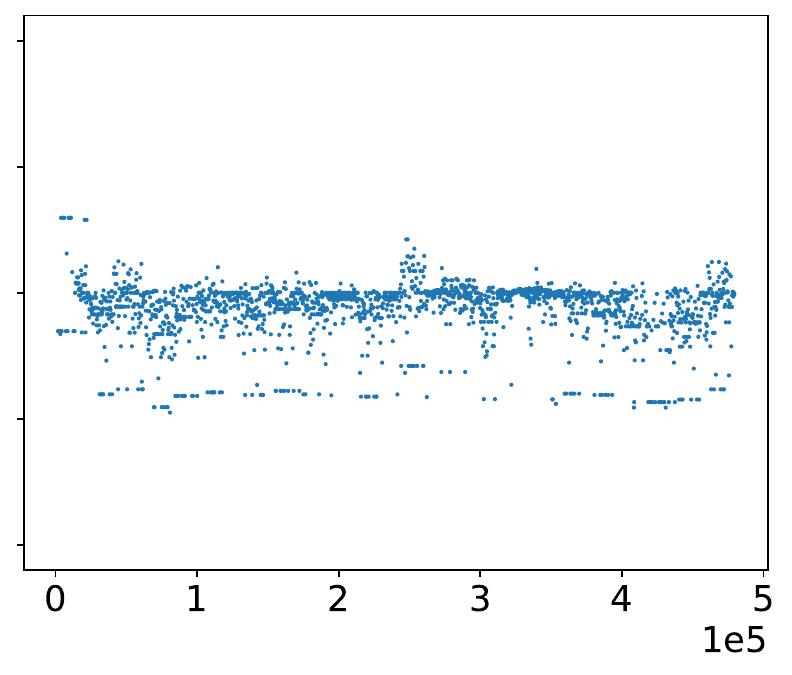}}
\subfloat[CVE-2016-4492]{\includegraphics[width=0.16\linewidth]{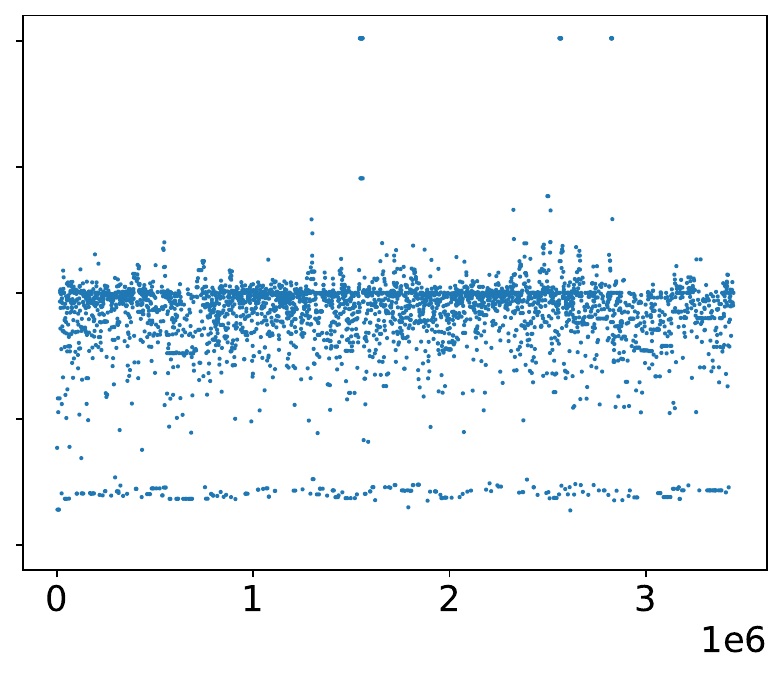}}
\subfloat[CVE-2018-8807]{\includegraphics[width=0.16\linewidth]{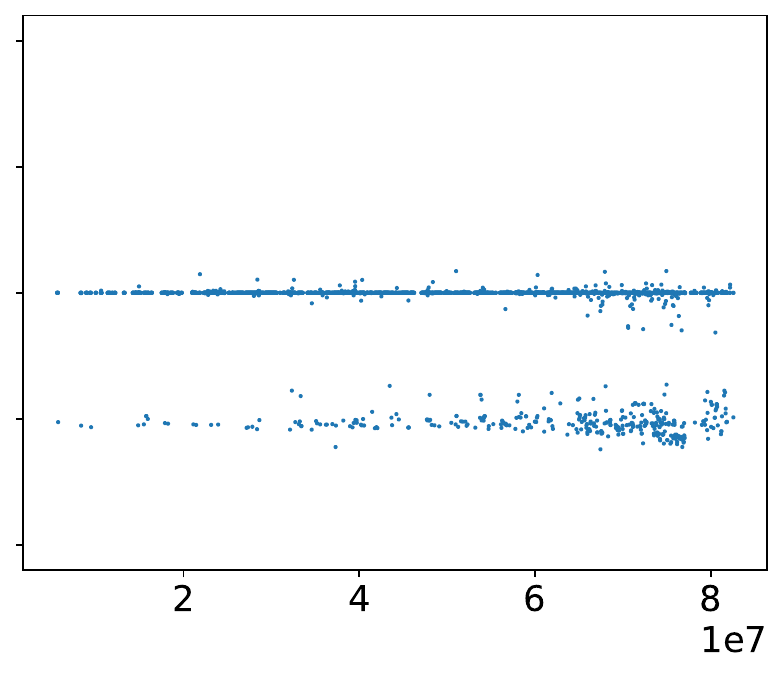}}
\subfloat[CVE-2017-9047]{\includegraphics[width=0.16\linewidth]{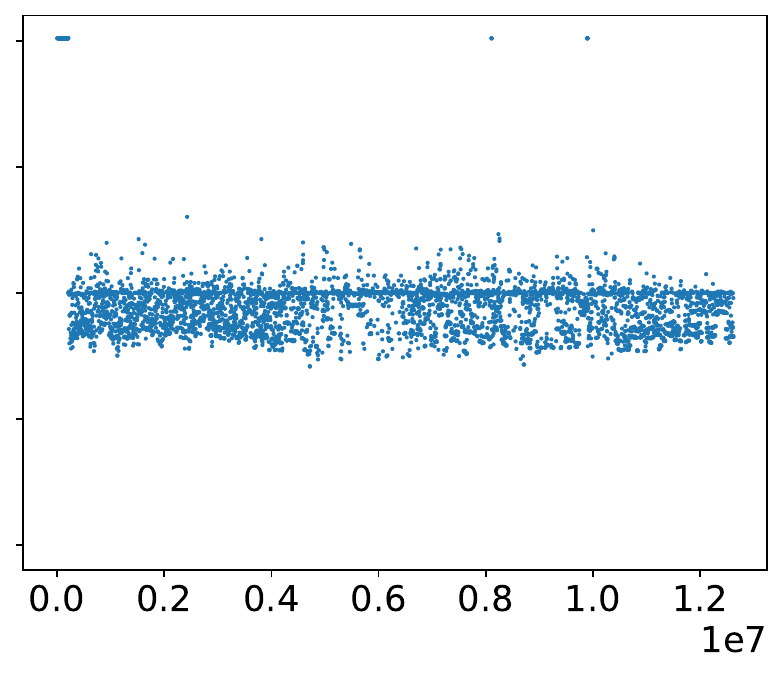}}
\caption{Distance decrease distribution on different CVEs in the benchmark. The y-axis is the distance decrease while the x-axis is the time (ms) the seed mutates to the testcase.}
\label{fig:scatter-decrease-by-time}
\end{figure*}

\begin{figure*}[!t]
\centering
\captionsetup[subfloat]{labelformat=empty}
\subfloat[CVE-2016-4487]{\includegraphics[width=0.177\linewidth]{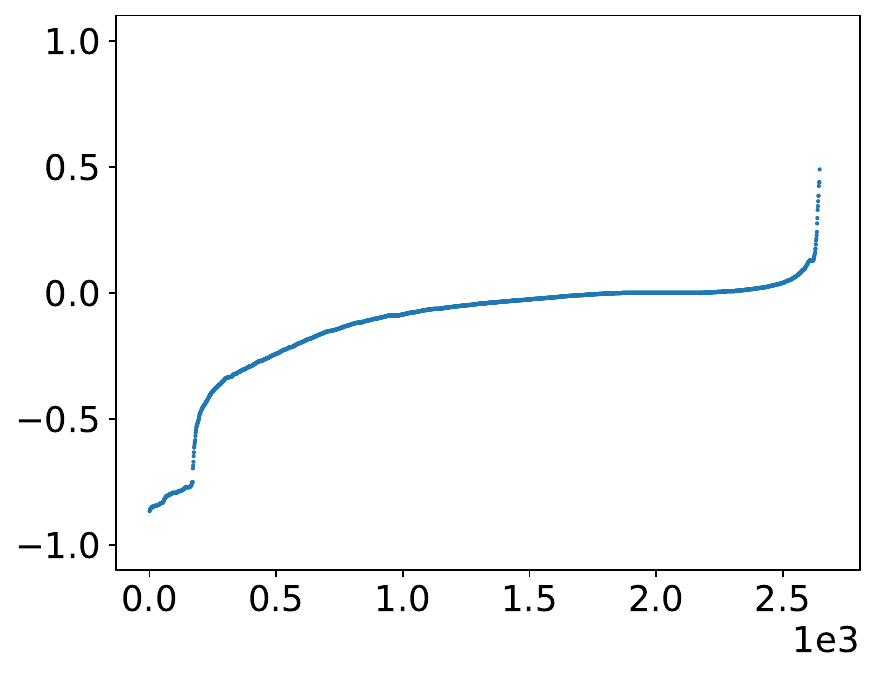}}
\subfloat[CVE-2016-4489]{\includegraphics[width=0.16\linewidth]{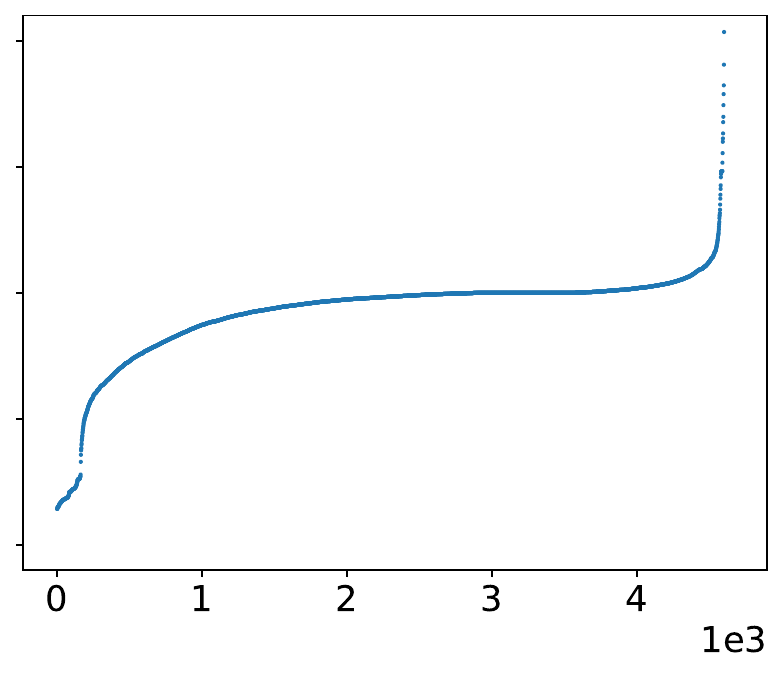}}
\subfloat[CVE-2016-4490]{\includegraphics[width=0.16\linewidth]{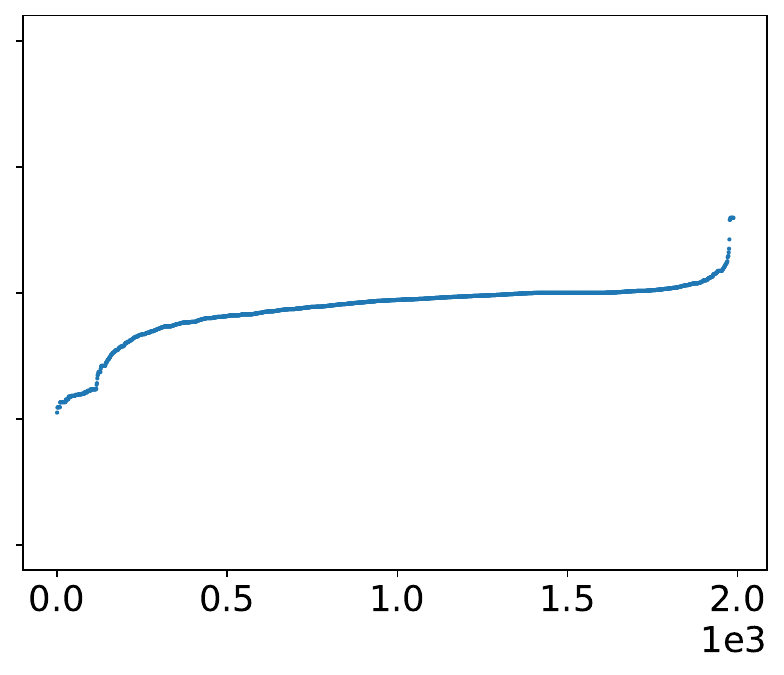}}
\subfloat[CVE-2016-4492]{\includegraphics[width=0.16\linewidth]{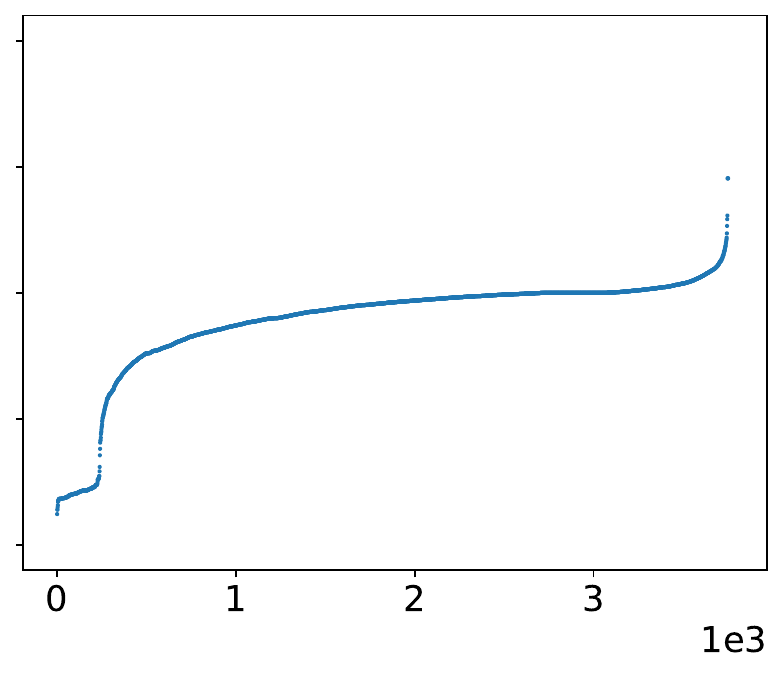}}
\subfloat[CVE-2018-8807]{\includegraphics[width=0.16\linewidth]{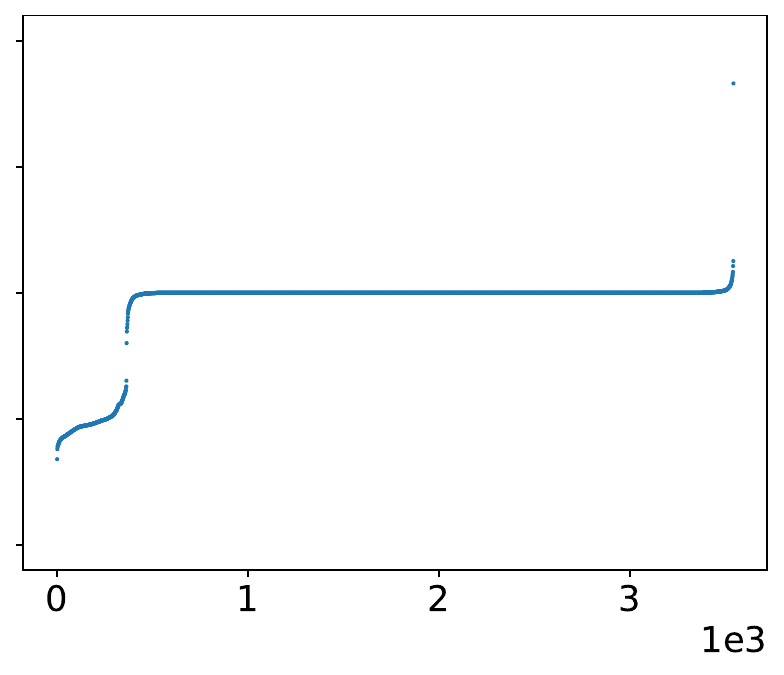}}
\subfloat[CVE-2017-9047]{\includegraphics[width=0.16\linewidth]{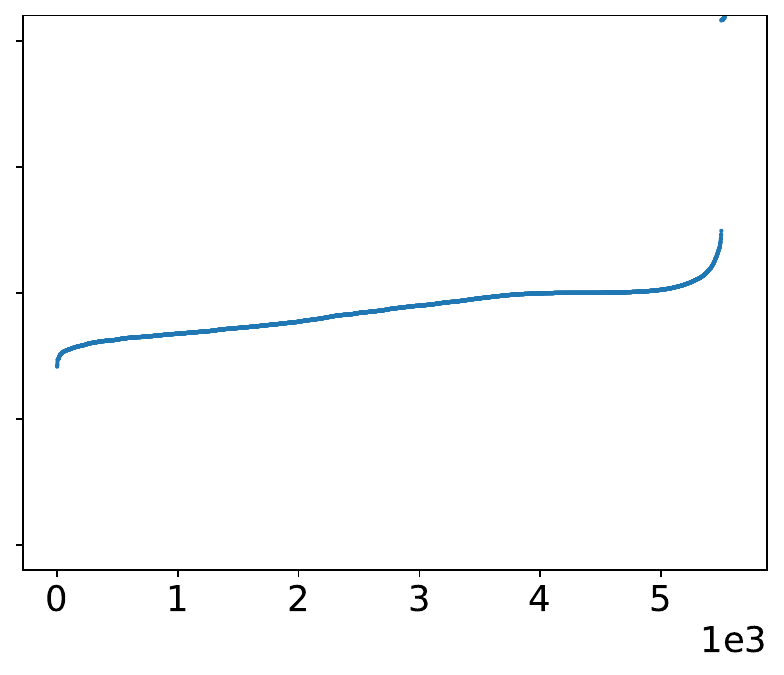}}
\caption{The cactus plot of distance decrease on different CVEs in the benchmark. The y-axis is the distance decrease while the x-axis is the number of mutation operations.}
\label{fig:decrease-cactusplot}
\end{figure*}

To further investigate the details of the evolution from the initial seed to the final PoC, we plotted line graphs to illustrate the variation of seed distance across all generations.

Specifically, for each PoC, we trace back to find all its ancestor seeds, recording when they are added to the queue and their distances during execution. 
Then, we plotted the variation of distance in each generation over time. 
Since the initial seed is often set to empty, its distance is set to a default large value.

As shown in Fig.~\ref{fig:lineage_of_PoC}, each red 'x' represents an ancestor seed of the PoC, with the rightmost red 'x' corresponding to the PoC itself. 
Overall, the distance of PoC significantly decreases relative to that of the initial seed. 
However, the distance of seeds does not always decrease gradually along the evolution. 
It is more common for the distance of seeds to fluctuate along all generations rather than consistently decrease compared to the previous generation. 

\textcolor{black}{
Moreover, it is worth noting that the final PoC seldom derives from the seed with the closest distance.
The seeds with further distance have more chances to generate the PoC.
This suggests that distance metrics based on the call graph and control flow graph have deviated from revealing the difficulty of approaching and triggering the target vulnerability, and encounter a bottleneck in selecting high-quality seeds and guiding directed fuzzing.
}

\begin{tcolorbox}[arc=2mm,title = {Lesson 2}]
For RQ3, the closest seed is not superior to other seeds in generating the PoC. On the contrary, the PoC often derives from the seeds with a slightly further distance. The distance metrics encounter a bottleneck in selecting high-quality seeds and guiding directed fuzzing due to its deviation from describing the difficulty of approaching and triggering the target vulnerability.
\end{tcolorbox}

\subsubsection{Mutation Struggles to Generate Closer Testcases}\label{sec:mutationstruggles}

According to the design philosophy of AFLGo, it iteratively selects a seed from the queue and calculates the total number of testcases (power scheduling) to be generated.
Typically, the amount of energy is mainly determined by seed distance and other relevant attributes (length, execution time, coverage, etc.). 
Then a large number of testcases are generated based on a certain mutation strategy and executed by the target program. 
However, AFLGo and many other existing directed fuzzers adopt the default mutation strategy of AFL, which is not sensitive to the target vulnerability or the target code location. 
This may hinder the fuzzer from generating testcases that progressively approach the target code location.

As shown in Fig.~\ref{fig:lineage_of_PoC}, we collect all testcases derived from all ancestor seeds of the PoC, including their distance and generation time. 
Although every ancestor seed of the PoC generates a large number of testcases (only a subset is displayed in the figure), a considerable portion of the descendant testcases have larger distances compared to their parent seed. 
This suggests that the existing mutation strategy may struggle to generate high-quality testcases with shorter distances.

We perform a quantitative analysis of the capability of the existing mutation strategy to generate high-quality testcases. 
Specifically, we assess the ability of mutation strategy to generate closer testcases by \textit{Decrease}, which is defined as:

\begin{equation}\label{eq:distance_decrease}
Decrease = \frac{Distance(T) - Distance(S)}{Distance(S)}
\end{equation}

where the testcase $T$ is directly derived from seed $S$. 

If the \textit{Decrease} exhibits an extremely small negative value, it implies that the distance of the descendant testcase is significantly reduced relative to its parent seed, thereby increasing the likelihood of approaching and triggering the target vulnerability. 
Conversely, a large positive value of \textit{Decrease} indicates that the descendant testcases have significantly greater distances compared to their parent seeds, which is an unfavorable outcome. 
In the scenario where \textit{Decrease} is close to 0, it is interpreted as only a negligible distance difference between the descendant testcase and its parent seed.

We plot the trend of \textit{Decrease} variation over time, as shown in Fig.~\ref{fig:scatter-decrease-by-time}. 
Each dot in the figure corresponds to the \textit{Decrease} of a testcase and its parent seed. 
Most of the dots are under the $x$ axis but very close to 0, implying mutation strategy yields a very limited distance decrease between the testcase and its parent seed. The corresponding dots of which $Decrease>1$ are not plotted, as they are exceedingly rare.

Furthermore, we have drawn the cactus plot of distance \textit{Decrease} as Fig.~\ref{fig:decrease-cactusplot}. 
Although the distance of most testcases decreased compared to their parent seeds, the decrease proportion is negligible. 
It is revealed that the overall distance \textit{Decrease} of all testcases relative to their parent seeds is only -13\% and the median of distance \textit{Decrease} is -5\%, suggesting that the distances of the majority of testcases remain nearly unchanged compared to their parent seeds.

Although it is a positive outcome that the highly randomized mutation strategy still yields an overall decreasing trend in the distances of testcases, it also indicates that there is room for improvement in the mutation strategy for directed fuzzing. 
If a mutation strategy tailored for directed fuzzing is designed, the overall efficiency of directed fuzzing could be further enhanced.

\begin{tcolorbox}[arc=2mm,title = {Lesson 3}]
Given the inefficiency of the current mutation strategy in generating high-quality testcases, we argue that there is room to improve the vanilla mutation strategy.
A dedicated mutation strategy tailored for the target code could significantly enhance the quality of seeds and the efficiency of directed fuzzing.
\end{tcolorbox}

\section{Discussion and Future Research}


In this section, we delve into domains for potential enhancement and further exploration in future research.

\subsection{From Distance to Reachability}

Measuring the quality of seeds is crucial in fuzzing. 
Existing distance metrics, as discussed in section \ref{sec:poc_lineage_analysis}, encounter a bottleneck in effectively selecting high-quality seeds and guiding directed fuzzing because it deviates from describing the difficulty of approaching and triggering the target code. 
This issue arises primarily because these distance metrics seldom consider the constraints along the potential path to the target code. 
It may be addressed by adding new information to the distance metric or creating a new metric to measure seeds. 
CAFL~\cite{CAFL} introduces the concept of data distance, which measures the difference between the values of relevant variables after execution and the conditions that these variables should satisfy to reach the target code location. 
CAFL calculates the data distances for all relevant variables and combines them with the code distance to construct a new distance metric that guides the DGF. 
Consequently, this new metric provides a more comprehensive indication of a seed's reachability to the target code, beyond simply measuring the code distance of a seed.

\subsection{Directed Mutation Strategy}

As discussed in section \ref{sec:mutationstruggles}, existing mutation strategies struggle to generate testcases closer to the target code due to its randomness. 
Therefore, reducing the randomness of existing mutation strategies may be a promising way to implement a directed mutation strategy.
For example, applying appropriate mutation operators on the critical bytes of the seed may contribute to generating more testcases closer to the target code.
MOPT \cite{mopt} performs an adaptive mutation strategy by scheduling the distribution of different mutation operators. 
Generally, the probability of using a specific mutation operator will be reduced if it contributes little to generate good testcases \cite{mopt}. 
The coarse-grained operators (mixed havoc, semantic mutation, splice) will be given less chance when the seed can reach the target function \cite{hawkeye}. 
GREYONE \cite{greyone} determines which input byte to mutate by analyzing the data dependency among input bytes, variables, and branches. 
These enhancements could serve as a foundation for research into directed mutation strategies, even though they were originally proposed to achieve higher code coverage.

\section{Related Work}

\subsection{Directed Fuzzing}

\textbf{Directed White-box Fuzzing.} The directed white-box fuzzing mainly uses symbolic execution engines to generate inputs that can efficiently explore the state space of feasible paths~\cite{klee, katch, bugredux}.
KLEE~\cite{klee} is capable of automatically generating high-coverage testcases for programs that have intensive interactions with the environment.
BuzzFuzz~\cite{buzzfuzz} utilizes dynamic taint tracking to locate critical areas within the original seed input file and generates new test input files through fuzzing operations.
KATCH~\cite{katch} combines symbolic execution with novel heuristic approaches based on static and dynamic program analysis, enabling fast testing of patch code and bug discovery upon patch introduction.
Dowser~\cite{dowser} proposes a guided targeted symbolic execution strategy to explore overflows in the code, rather than covering all code paths.
Additionally, Directed Symbolic Execution (DSE)~\cite{ma2011dse}, which utilizes symbolic execution to reach specific target locations within programs, is suitable for many other tasks such as patch testing~\cite{santelices2008testaugmentation, bohme2013partition}, reaching rare code~\cite{xu2010directedtrade}, and reproducing bug~\cite{bugredux}.

However, since symbolic execution relies on heavy-weight program analysis and constraint solving, existing directed white-box fuzzing suffers from scalability and compatibility limitations \cite{Wang2020SoKTP}. 
When faced with large software, directed white-box fuzzing requires enormous computing resources and has difficulty in generating sufficient inputs that can reach the target code in a reasonable time.

To address this problem, the concept of hybrid fuzzing is proposed by combining directed fuzzing with concolic execution and directed fuzzing.
Driller~\cite{stephens2016driller} uses inexpensive fuzzing to exercise compartments of an application and uses concolic execution to generate inputs that satisfy the complex checks separating the compartments. The strengths of the two techniques are combined while their weaknesses are mitigated.
SDHF~\cite{seqhybridfuzz} combines directed grey-box fuzzing and concolic execution and uses user-specified statement sequences to guide fuzzing
SAVIOR~\cite{chen2020savior} enhances hybrid testing with bug-driven prioritization and bug-guided verification.

\textbf{Directed Grey-box Fuzzing}. 

The philosophy underlying DGF is to design a fitness function to measure the capability of a seed to approach and trigger the target code.
AFLGo~\cite{AFLGo} is the first DGF that measures seed mainly based on the distance between the execution trace of the seed and the target code. 
Seeds with shorter distances will be assigned more energy and generate more testcases based on them. 
As a consequence, the probability of triggering the target code increases.

Subsequent advancements in DGF have introduced various strategies to optimize distance calculation methods
Hawkeye~\cite{hawkeye} measures the distance between functions based on the pattern of the immediate call relation. 
CAFL~\cite{CAFL} proposes the concept of data distance to measure the seeds from the perspective of satisfying the data constraints. 
WindRanger~\cite{Windranger} introduces Deviation Basic Blocks (DBBs) to identify basic blocks whose execution may deviate from the direction towards the target code. 
It argues that not all basic blocks in the execution trace are equally important and the coverage of DBBs indicates the capability of a testcase to reach the target code. 
DAFL~\cite{DAFL} identifies target-related basic blocks based on the Def-Use Graph. 
Only the basic blocks in these relevant functions are instrumented and contribute to the distance calculation.

Directed grey-box fuzzing has demonstrated notable effectiveness for specialized tasks. 
UAFuzz~\cite{UAFuzz} adjusts distance weights based on the capability to trigger allocation, use, and free events sequentially, and proposes the first fuzzing engine tailored to UAF vulnerabilities.
1dFuzz~\cite{1dfuzz} uses the trailing call sequence (TCS), the common and unique feature of patches, to locate the target code location and present a novel directed differential fuzzing solution to reproduce 1-day vulnerabilities. 
FairFuzz~\cite{fairfuzz} proposes a novel lightweight mutation masking strategy to increase the chance of hitting the program regions that are missed by previously generated inputs.
AFLChurn~\cite{aflchurn} performs regression testing in the DGF way by targeting code that has changed recently or often.

Several studies have improved seed quality through path pruning and seed selection strategies, effectively guiding directed grey-box fuzzing and reducing unnecessary overhead. 
FuzzGuard~\cite{fuzzguard} trains a deep neural network to predict the reachability and filter out unreachable testcases, which significantly reduces the time spent on real execution.
Beacon~\cite{beacon} proposes a lightweight static analysis method to obtain the weakest precondition of exposing the target vulnerability and then prune off infeasible paths. 
MC2~\cite{mc2} defines the directed grey-box fuzzing as an oracle-guided search problem and employs execution complexity to measure a fuzzing algorithm’s asymptotic performance.
FishFuzz~\cite{fishfuzz} proposes a novel seed selection strategy that suits multiple-target scenarios and dynamically prioritizes the valuable targets.

Static analysis techniques can enhance seed quality by leveraging crucial information.
TIFF~\cite{tiff} infers the type of the input bytes by in-memory data-structure identification and dynamic taint analysis. 
It maximizes the likelihood of triggering memory-corruption bugs by generating fewer but better inputs.
SemFuzz~\cite{semfuzz} uses the NLP technique to extract vulnerability-related information (e.g., CVE reports and Linux git logs) and proposes a semantics-based fuzzing process to guide the automatic generation of PoC exploits.

\subsection{Directed Fuzzer Evaluation}

Due to the prosperity of fuzzing, researchers conduct surveys and evaluations on the design philosophy and effectiveness of fuzzers \cite{unifuzz, klees18evalutefuzztesting, Wang2020SoKTP}. 
Wang et al.~\cite{Wang2020SoKTP} categorize existing fuzzers by the fuzzing methodology and discussed the challenges and future trends. 
Li et al.~\cite{unifuzz} propose an open-source and metrics-driven platform to assess fuzzers, calling for a comprehensive and quantitative evaluation manner to avoid unilateral conclusions. 

As for the directed fuzzing evaluation, Lee et al.~\cite{ontheeffectiveness} discuss the overfitting problem in directed grey-box fuzzers and use Fuzzle, a state-of-the-art bug synthesis tool, to build benchmarks and avoid the uneven difficulty of CVEs in real-world software. 
Kim et al.~\cite{fse24areweheadinrightdirection} reveal the potential pitfalls (target bug specification and crash triage method) and proposed guidelines to evaluate fuzzers with the appropriate statistical test. 
Meanwhile, our work provides valuable insights into the distance metric and mutation strategy, which are the most important components in DGF but are overlooked by previous works.

\section*{Acknowledgments}
We sincerely appreciate the reviewers for their valuable comments to improve our paper.
This work is supported by the NSFC under No. 62202484.

\section{Conclusion}

In this paper, we conduct the first empirical study on the distance metrics in guiding directed grey-box fuzzing.
We systematically discuss the typical distance metrics in terms of calculation methods and granularities.
Then, we implement different distance metrics based on the same framework (i.e., AFLGo).
We conduct comprehensive experiments to evaluate the performance of each distance metric in triggering the target code.
From the experimental results, we can find the following insights.
In terms of distance calculation methods, arithmetic mean distance has a slight advantage over harmonic mean and closest path distance.
In terms of distance granularity, function-level distance performs the worst compared with approximated basic-block-level distance and basic-block-level distance.
However, the difference among different distance metrics is not very significant.
Moreover, we explore the capability of closer seeds in generating testcases with closer distances.
We find that no more than 13\% of testcases have a closer distance than their parent seed.
Based on in-depth analysis, we find that the reason may be the inefficiency of the existing DGF mutation strategy.
Despite the selection of high-quality seeds, the current mutation strategies are too random to utilize these seeds fully.
This calls for the need to design effective mutation strategies. 
Therefore, it is essential to design effective mutation strategies tailored for DGF.
The code and dataset in our paper are available at \href{https://github.com/slient2009/DistanceMeasurement}{https://github.com/slient2009/DistanceMeasurement}.
We hope our study could inspire future research in DGF.

\bibliographystyle{ieeetr} 
\bibliography{main}

\end{document}